\documentstyle[preprint,aps,prl,epsf]{revtex}
\begin{document}
\def\bea{\begin{eqnarray}}
\def\eea{\end{eqnarray}}
\def\a{\alpha}
\def\d{\delta}
\def\p{\partial} 
\def\nn{\nonumber}
\def\r{\rho}
\def\rv{\bar{r}}
\def\la{\langle}
\def\ra{\rangle}
\def\e{\epsilon}
\def\o{\omega}
\def\n{\eta}
\def\g{\gamma}
\def\break#1{\pagebreak \vspace*{#1}}
\def\f{\frac}
\title{Surface tension and the cosmological constant${}^*$}
\author{Joseph Samuel and Supurna Sinha}
\address{Raman Research Institute,
Bangalore, India 560080\\}
\date{${}^*$Dedicated to Rafael Sorkin on his sixtieth birthday}
\maketitle
\widetext
\begin{abstract}
The astronomically observed value of the cosmological 
constant $\lambda$  is small but non-zero. This 
raises two questions together known as the
cosmological constant problem a) why 
is $\lambda$ so nearly
zero? b) why is $\lambda$ not {\it exactly} 
zero? 
Sorkin has proposed that b) can be naturally
explained as a $1/\sqrt{N}$ fluctuation by invoking 
discreteness of spacetime at the Planck scale due to quantum gravity. 
In this paper we shed light on these questions by developing an 
analogy between the cosmological
constant and the surface tension of membranes.
The ``cosmological constant  problem'' has a natural
analogue in the membrane context: the vanishingly small surface tension
of fluid membranes  
provides an example where question a) above arises
and is answered.
We go on to find a direct analogue of Sorkin's proposal for answering
question b) in the membrane context, where the discreteness of spacetime
translates into the molecular structure of matter. 
We propose analogue experiments to probe a small and fluctuating
surface tension in fluid membranes. A counterpart of dimensional
reduction a la Kaluza-Klein  and large extra dimensions also appears in 
the physics of fluid membranes.
\end{abstract}

\pacs{PACS numbers: 04.60.-m, 82.39.Wj, 87.15.Kg, 87.16.Dg, 46.70.Hg,
68.03.-g }                                         

\narrowtext
{\it Introduction}:
Condensed matter analogues have been vital to progress
in fundamental physics.
Ferromagnets and superconductors have led to ideas like spontaneous
symmetry breaking and the
Higgs phenomenon.
More recently, laboratory analogues
\cite{unruhana,visser}
of Hawking radiation have been discussed
in the context of supersonic fluid flows and Bose-Einstein
condensates. Analogue gravity in superfluid Helium
is being vigourously pursued\cite{volovik}. Another
example is the analogy \cite{ajit,vilenkin}
between defects in liquid crystals and phase transitions
in the early universe. There has also been work \cite{capo}
exploring parallels between the differential geometry 
of soft condensed matter systems and general relativity.  
In some domains of fundamental physics like gravity at the Planck scale,
experiments cannot be performed because they are beyond our reach in 
energy. 
Laboratory analogues are therefore extremely valuable as they
provide a concrete experimental context for ideas in 
fundamental physics and
are the
nearest one can get to ``experimental quantum gravity''. 
Experiments, whether real or gedanken, enrich the field by sharpening
questions.

The purpose of this paper is to develop and explore an analogy between the
cosmological constant $\lambda$
and the surface tension $\sigma$ of membranes. We first describe the
cosmological constant problem in general terms and then show using
our analogy that similar problems appear in soft condensed matter physics.
The analogy is fairly good and we are able to translate ideas from one context
to the other. We develop the analogy in detail, discuss the insights
gained from it and conclude with a discussion of the limitations of the
analogy.

{\it Cosmological Constant problem:}
Let us briefly recount the cosmological constant problem. 
In general relativity (GR), a spacetime is a pair $({\cal M},g)$, 
where ${\cal M}$ is a four dimensional manifold and $g$ a Lorentzian 
metric.
We will sometimes refer to $({\cal M},g)$ as a {\it history} ${\cal H}$.
The dynamics of GR is described 
by the Einstein Hilbert action $I_2=c_2\int d^4 x \sqrt{-g} R$,
modified by the addition of a cosmological 
term $I_0=c_0\int d^4 x \sqrt{-g} $. To connect with standard 
notation, $c_2=1/(16\pi G)$, where $G$ is Newton's constant and $c_0$ 
is conventionally referred to as $\lambda$.
Usually, higher derivative 
terms like $I_4=c_4\int d^4 x \sqrt{-g} R^2$ are dropped as being 
negligible. This is entirely
in the spirit of effective field theory (or Landau theory in condensed
matter physics), where we expect that the low energy physics will 
be dominated by the lower derivative terms. However, applying this logic
to general relativity, we would expect the cosmological constant term 
$I_0$ to dominate over the Einstein-Hilbert term $I_2$. 
A crude dimensional analysis would suggest a value for the 
cosmological constant which is of order $1$ in dimensionless 
units ($G=c=\hbar=1$). In fact, the observed value of the cosmological 
constant 
is practically zero. But not exactly zero! 
The Planck length $l_{Planck}=10^{-33}cm$ serves as a natural unit of 
length in this problem and astronomical observations give $\lambda 
l_{Planck}^4=10^{-120}$:
tiny but non-zero.

The dilemma of the cosmological constant
thus has two horns\cite{rafael}:
\begin{description}
\item a) Why is the cosmological constant nearly zero?
\item b) Why is it not {\it exactly} zero?
\end{description}
It seems hard to come up with a natural explanation for {\it both} these facts: 
one could conceivably construct models (for example, by invoking
a symmetry) in which $\lambda$ exactly
vanishes. But why then does it only approximately 
vanish\cite{gulliver} in the real world? There is now compelling 
astrophysical evidence\cite{evidence} for a small and nonzero $\lambda$. 
Many efforts to understand this observation suffer from some kind
of ``fine tuning problem'' and are therefore not a natural \cite{quint}
explanation.

{\it Sorkin's proposal:}
An exception is the beautiful idea due to 
Sorkin \cite{rafael} that quantum gravity may provide a natural 
explanation 
stemming from a fundamental discreteness of spacetime at the Planck scale.
Sorkin's proposal for solving the cosmological constant problem 
is in the framework of causal sets. It is widely acknowledged that
though the smooth manifold model of spacetime works well over a range of 
length scales, such a picture
may not hold at the Planck scale. (See remarks by Einstein
quoted in \cite{gomb}).
The finiteness of black hole entropy induces\cite{large} us to believe 
that
spacetime is discrete at this scale.
In Sorkin's approach of causal set theory,
one replaces spacetime by a discrete structure, a collection of points 
carrying causal relations. The number $N$ of points is associated with the 
total four-volume ${\cal V}=\int d^4 x \sqrt{-g}= l_{Planck}^{4} N$ of 
spacetime. 
The rest of the metrical information of spacetime (the conformal 
structure) is captured in causal relations between points.  
Spacetime is regarded as an emergent notion,
when the number of points $N$ gets large. 

The spacetime four volume ${\cal V}$ also plays a role in unimodular 
gravity\cite{einstein,weinberg},
a slight modification of GR, which has been studied with the 
hope of solving the problem of ``time"\cite{unruh} in quantum gravity. 
When one varies the action in unimodular gravity,
one only allows variations that
preserve the unimodularity of the metric. This theory 
is classically equivalent to GR modified by a cosmological constant.
GR and unimodular gravity
have identical predictions for solar system physics (since $\lambda$ is
negligible at this scale) and so unimodular
gravity shares the experimental success of GR.
However, unlike in GR, the 
cosmological constant $\lambda$ is not a coupling constant, but a Lagrange
multiplier which enforces unimodularity of the metric.

Sorkin's proposal 
addresses horn $2$ of the cosmological constant dilemma. Let us for the 
moment suppose that horn $1$ has been solved: some mechanism has been 
found for ensuring that the cosmological constant is zero. Sorkin's idea
is that there will be fluctuations about this mean value which result in 
a small nonzero cosmological constant. These fluctuations have their 
origin in quantum gravity. The order of magnitude of these fluctuations is 
$\frac{1}{\sqrt N}$, where $N$ is the four volume of the universe 
expressed in Planck units. In unimodular gravity the cosmological constant 
$\lambda$ is a dynamical variable and conjugate to 
the four volume ${\cal V}=l_{Planck}^4 N$ of spacetime.
Sorkin proposes $1/\sqrt{N}$ fluctuations in $\lambda$
as the mechanism for a small
cosmological constant. Based on the argument given, Sorkin
predicted\cite{rafael} a value for the cosmological
constant which is of order
\begin{equation}
\label{lambda}
\lambda\approx \frac{l_{Planck}^{-4}}{\sqrt{N}}
\end{equation}
and of fluctuating sign. In this model, the root mean squared
fluctuation in the vacuum 
energy density ($\lambda$)
is comparable in magnitude to the matter density at {\it all} epochs.

These predictions are consistent \cite{evidence} with
astronomical data (redshift-luminosity distance relations)
from type I supernovae:
the observations show that the
universe is accelerating at the present epoch,
indicative of a positive cosmological constant.
Sorkin's argument predicts the correct
order of magnitude for
$\lambda$. 
Other researchers \cite{paddy,volovik2}
have since taken up this idea with slight variations. In this paper
we have followed Sorkin's original proposal\cite{rafael} and treatment.

{\it  Membranes in soft matter physics:}
Let us now turn from GR and the cosmological constant to membranes
in soft condensed matter physics. 
A configuration ${\cal C}$ of a membrane 
is described as a $2$-dimensional surface
$\Sigma$ embedded in ordinary flat three dimensional space. $\Sigma$ 
inherits a metric $\gamma$ from this space.
This permits us to define an area
element $d^2x \sqrt{\gamma}$ on $\Sigma$. The embedding also
determines two curvatures: the extrinsic curvature $H$ and the
intrinsic curvature $K$ of $\Sigma$. Note that $H$ has dimension
$1/L$ of inverse length, while $K$ has dimension $1/L^2$. To complete
this description of a membrane, we need to specify the energy
of a configuration ${\cal E}$. We restrict ourselves to membranes
which have two ``sides'' (orientable) and which are symmetric in their
two ``sides''.  The latter implies that the energy is invariant under 
$H\rightarrow -H$.
In the spirit of Landau theory, we write down 
terms with the lowest
number of derivatives consistent with the symmetry of the 
problem\cite{piran}. 
\begin{eqnarray*}
{\cal E}_0=a_0 \int_{\Sigma}d^2x \sqrt{\gamma}\\
{\cal E}_2=a_2 \int_{\Sigma}d^2x \sqrt{\gamma} H^2+a'_2 
\int_{\Sigma}d^2x \sqrt{\gamma} K
\end{eqnarray*}
The leading term here is the surface tension $a_0$, which is 
conventionally denoted as $\sigma$.
Higher derivative terms like
${\cal E}_4=a_4 \int_{\Sigma}d^2x \sqrt{\gamma} H^4+a'_4 
\int_{\Sigma}d^2x \sqrt{\gamma} K^2+a''_4 \int_{\Sigma}d^2x \sqrt{\gamma} 
H^2 K$
are negligible in the long wavelength description. The physics of 
membranes
is then contained in the partition function 
\begin{equation}
Z=\Sigma_{\cal C} \exp{-{\cal E}({\cal C})/(k_BT)}
\label{partition}
\end{equation}
where ${\cal E}={\cal E}_0+{\cal 
E}_2+...$ is an expansion of the energy in inverse powers of length.
Henceforth we will set Boltzmann's constant $k_B$ to unity and measure
temperature in energy units.
As we explain in greater detail below, this mathematical
model of a membrane can be physically realised as 
an interface between fluids. 

{\it Analogy between Membranes and Spacetime:}
As the reader will readily appreciate, there is a clear analogy
between the GR situation and the soft matter one. The analogy
is based on the usual correspondence between quantum physics
and statistical mechanics. A {\it history} ${\cal H}$ in GR is 
replaced by a {\it configuration} ${\cal C}$ in statistical physics.
A sum over histories in quantum GR $\Sigma_{\cal H}\exp{i I({\cal 
H})/\hbar}$ is replaced 
by a sum over configurations (\ref{partition}) with Boltzmann weight.
The {\it action} $I({\cal H})$ is replaced by the {\it energy}
${\cal E}({\cal C})$. The expansion of the action in powers
of increasing mass dimension is similar to the Landau theory
expansion of the energy in powers of decreasing length dimension.
Just as quantum effects lead to fluctuations about the classical 
path of least action (the history solving the classical equation),
thermal effects cause fluctuations about the minimum energy configuration.
The role of Planck's constant is played
by the temperature $T$. The leading term in the action
is the cosmological constant term just as the leading term 
in the energy of a membrane is the surface tension term. 
The surface tension has the interpretation of ``energy cost per unit area 
of membrane'': one has to supply energy to increase the area of 
the membrane. This is 
usually supplied in the form of mechanical work when one works up a lather 
while shampooing or beating an egg.
In GR, the cosmological constant is the ``action cost per
unit four volume of spacetime''.
The analogy is summarised in table 1 for ready reference.

The geometric description of a membrane as a smooth two manifold 
$\Sigma$ embedded in space is only a mathematical idealisation. 
A real membrane in the laboratory is composed of molecules.
The smooth manifold picture of ${\Sigma}$ is only 
valid at length scales large compared to the molecular length scale
$l_{mol}$. This is quite similar to the breakdown of the 
smooth manifold picture of spacetime at the Planck scale.
The role of the Planck length is played here
by the mean intermolecular spacing $l_{mol}$, 
which is about $0.3 nm$\cite{daoud}.                
At mesoscopic scales, the membrane appears
smooth and in a statistical sense, locally homogenous and isotropic.
For instance the probability of having
a void of area ${\cal A}_{void}$
in a membrane of area ${\cal A}$
can be crudely estimated as
$P_{void}\approx{\cal A}/{\cal A}_{void} \exp{-{{\cal 
A}_{void}/l_{mol}^2}}$.
This works out to $P_{void}\approx   {\cal A}/{\cal A}_{void} \exp{-10^7}$,
for a micron sized void.
This is similar in spirit to estimates in the causet framework
for the probability of
nuclear sized voids in the age of the universe:
$P_{void}\sim e^{-10^{80}}$\cite{fay}.

Using the analogy, we would expect that the surface tension
of a membrane $\sigma$ is of order $1$ in dimensionless units, 
that is $\sigma\approx T/l^2_{mol}$.
It is instructive to see what happens if one starts with a microscopic
energy in which the surface tension is set to zero {\it by hand}. 
Consider such a membrane whose  equilibrium configuration is a plane 
rectangle with sides $L_1,L_2$ and area ${\cal A}=L_1 L_2$.
Due to thermal fluctuations, the membrane will vibrate about its
equilibrium configuration. Assuming small vibrations, we can model
them as harmonic oscillators and expect by equipartition that 
the expectation value of energy $<E>$ in each mode is $T$. 
Performing a sum over modes to evaluate
the contribution from all the modes we find a divergent answer 
which has to be regulated by the molecular scale cutoff.
\begin{equation}
T\int\int_0^{k_{max}} \frac{d^2k d^2x}{(2\pi)^2}
\label{modesum}
\end{equation}
where $k$ is a wave vector and $k_{max}=2\pi/l_{mol}$ is the
cutoff set by the molecular scale. Performing the $k$ integral we find
that this contributes a term
 \begin{equation}
\frac{\pi T}{l^2_{mol}} \int d^2x =\frac{\pi T}{l^2_{mol}} {\cal A}
\label{radiative}
\end{equation}
to the energy, giving rise to a surface tension of order 1 in 
dimensionless units. Even if one assumes that the microscopic energy
has zero surface tension, such a term is generated by ``radiative 
corrections'' in a manner analogous to the generation of vacuum energy
from the Casimir effect.
The spontaneous generation of a surface tension by
radiative corrections can be viewed as a flow of coupling constants
in the renormalisation group sense. Unless protected by 
symmetry\cite{symm},
coupling constants will flow. 

We therefore expect that interfacial tensions will be of order
$T/l^2_{mol}$. 
Using the values $k_BT=1/40 eV$  (corresponding to $300^0 K$) 
and $l_{mol}=.3nm$,
we expect that the surface tension of membranes to be around 
$ 40$ in units of  milli Joules per square metre. This  
expectation turns out to be correct. Table 2 shows the surface tensions
for interfaces of simple liquids. 
They all have the expected order of magnitude \cite{daoud}. 
For example, the air-water interface has a surface tension 
of  $ 40 m J m^{-2} $. Adding soap to water
decreases the surface tension
by a factor of three but does not change its order of magnitude.
For some membranes involving complex
molecules, the $l_{mol}$ is higher \cite{thickness} by a factor of $30$ and the
surface tension is correspondingly lower by three orders of magnitude,
again consistent with the dimensional argument. Thus most membranes
do not suffer from the analogue of the cosmological constant problem.
This reinforces our faith in the dimensional argument.

{\it Fluid membranes and the cosmological constant problem:}
However, there is an exception to the dimensional
argument which is of great interest from the present perspective:
fluid membranes.
These are 
characterised by a negligibly small surface tension $\sigma$, {\it orders of
magnitude below that predicted by the dimensional argument}. 
The statistical mechanics of fluid membranes is dominated by the curvature
terms ${\cal E}_2$ rather than by the surface tension term ${\cal E}_0$. 
This is an exact counterpart of  
the fact that $I_2$ dominates over $I_0$
in GR. Fluid membranes thus provide us with an example in which
part a) of the cosmological constant problem 
is naturally solved. To see what we can learn from this let us consider
fluid membranes in more detail and understand why they have vanishing
surface tension. The review of fluid membranes  
and their vanishing surface tension is based on \cite{safran} to which
the reader is referred for more details.

{\it Fluid Membranes and their vanishing surface tension:} 
A fluid membrane\cite{safran,barrat} is a two-dimensional assembly of 
surfactants or amphiphilic molecules, consisting of hydrophilic 
(water loving) polar head groups and 
hydrophobic (water hating) hydrocarbon tails. If amphiphilic
molecules are added to oil-water mixtures, they prefer to stay at the 
interface so as to satisfy both parts of the molecule.
If amphiphiles are added to
a single solvent like water, beyond a critical concentration,
they tend to  form aggregates ({\it e.g.} micelles, vesicles)
with the hydrocarbon
tails tucked away from the water and the polar heads in contact with
the water. An example of such an aggregate is a bilayer.
These bilayers are clearly symmetric in their two sides. One
example (of great biological interest) of such fluid membranes
are lipid bilayers.

The solubility of amphiphilic 
molecules in water is low and molecules prefer to
stay on the membrane rather than dissolve in the bulk.
As one increases the volume fraction of amphiphiles the molecules  
pack more and more  densely in the membrane and the area per molecule
$\alpha$ decreases. There is a limit to this
packing density however, and at a critical value of $\alpha=\alpha_0$,
there is a minimum in the free energy per molecule $f(\alpha)$. A further
increase in the number of molecules does not decrease $\alpha$, but
increases the area of the interface (by rippling for example)
so as to accommodate the increase
of molecules. Such a membrane is said to be saturated. At the saturation
point $\alpha=\alpha_0$ the free energy per molecule has a minimum
\begin{equation}
\frac{\partial f}{\partial \alpha}|_{\alpha=\alpha_0}=0
\label{minimum}
\end{equation}

Consider a saturated membrane with a fixed area ${\cal A}$. The number
of molecules on the membrane satisfies ${\cal A}=N\alpha$. The total free
energy is given by $F({\cal A})=N f(\alpha)$. Computing the surface 
tension
of the membrane by differentiating $F({\cal A})$ with respect to ${\cal 
A}$ to 
find the expected value of the surface tension $\sigma$
we find that 
\begin{equation}
<\sigma>=\frac{\partial F}{\partial {\cal A}}=\frac{\partial f}{\partial 
\alpha}|_{\alpha_0}=0
\label{surfacetension}
\end{equation}

As a result the surface tension is no longer relevant to the problem
and the behaviour of the membrane is dictated by the higher derivative
terms in the free energy, the curvature terms. Forcibly stretching the
membrane would result in more molecules coming out of solution and
sticking to the membrane till the preferred value of area per molecule
is attained.
This implies that the interfacial tension associated with 
the membrane is effectively zero\cite{footren}. 
Thus we find that fluid membranes are a natural soft condensed 
matter example in which part a) of the cosmological constant
dilemma\cite{rafael} is resolved.  

 {\it Fluid Membranes and fluctuating surface tension:} 
Interestingly, the second horn of the dilemma can also be addressed in this
condensed matter context. A fluid membrane consists of a finite number
$N$ of discrete elements or molecules. Therefore, just like the 
fluctuations in the cosmological constant which appear in the 
discrete quantum gravity models \cite{rafael} a fluid membrane
consisting of a finite number $N$ of molecules has an
interfacial tension $\sigma$ which fluctuates about zero. The argument
just consists of differentiating the free energy $F({\cal A})$ once more 
with
respect to the area to calculate the 
mean square statistical fluctuations in the surface tension:
\begin{equation}
(\Delta \sigma)^2=<(\sigma-<\sigma>)^2>= T \frac{\partial^2 F}{\partial 
{\cal A}^2}=\frac{T}{N} \frac{\partial^2 f}{\partial 
\alpha^2}|_{\alpha_0}.$$
\label{ms}
\end{equation}

We can (naturally!) expect $T\frac{\partial^2 f}{\partial \alpha^2}|_{\alpha_0}$
to be of order $1$ in dimensionless units and therefore we find
\begin{equation}
\Delta \sigma\sim \frac{1}{\sqrt{N}}
\label{rms}
\end{equation}
in complete analogy to Sorkin's proposal in the cosmological context.

To summarize, thermal fluctuations of a finite number of molecules 
in the context of fluid membranes are analogous to quantal fluctuations 
associated with a finite number of discrete spacetime elements in the 
causet model of quantum gravity. These fluctuations are finite size
effects and dissappear ({\it i.e} $\Delta \sigma\rightarrow 0$) in 
the thermodynamic limit ($N \rightarrow \infty$) 
just as in the cosmological context 
at late epochs Sorkin's argument predicts that 
the cosmological constant fluctuations dissappear. 

{\it Experiments:}  The fluctuating surface tension of fluid
membranes can be experimentally probed. We briefly describe possible
experiments in idealised form. Consider a cylindrical fluid membrane
in an ambient buffer solution stretched between two tiny (say tens of 
nanometres) rings. The ends of the cylinder are open so that the pressure
on the two sides of the membrane are balanced. One of the rings is 
attached to a piezoelectric translation stage (figure) which can be moved in 
nanometre steps. The other ring is attached to a micron sized bead which 
is confined in an optical trap. If $r$ is the radius of the rings, 
the area of the membrane is given by $2\pi r L$, where $L$ is the separation
between the rings. At fixed separation $L$, one can measure the force $F$ 
on the micron sized bead by looking (through a microscope) at its 
displacement in the optical trap. This force $F$ is related to the 
surface tension by $F=2\pi r \sigma$. Thus, one can directly measure
the surface tension of a small fluid membrane as a function of its area.
We expect to find a small fluctuating component in the surface tension,
whose magnitude decreases as the inverse square root of the
area of the  membrane. In order for this $1/\sqrt{N}$ effect to be 
appreciable, $N$ has to be suitably small. For a $.1\%$ effect we
need $N=10^6$. Variants of this setup \cite{vesicle} can be considered, 
for example using 
a strip geometry for the membrane rather than a cylindrical one. One could 
also let the bead ``recoil'' by switching off the trap\cite{roopa}.

The experiments proposed above are well within the
realm of possibility. In fact similar experiments have
already been done (albeit with a completely different motivation).
In \cite{roopa}, the authors report an experiment done on vesicles
which is similar to the experiment we have proposed. They
pull out a tube ($80 nm$ radius) of lipid membrane
from a multilamellar vesicle                                      
of DDAB, stretch it out over tens of microns,
and measure its force extension relation. While \cite{roopa}
do measure a surface tension it is not
the effect that we discuss here.

{\it Kaluza-Klein compactification:}
One could view the circular cross section of the
cylinder as a `compactified' internal dimension.
The cylinder can be regarded as a line's worth of circles.
From a coarse perspective one can disregard the compact
circular dimension and view the cylinder as a line.
From dimensional reduction the
line tension gets a contribution from the
(extrinsic) curvature of the compactified dimension.
The expected value of the surface tension (in order of magnitude) is 
$\sigma=T/r^2$,
where $r$ is the radius of the compact dimension. 
The measured value \cite{roopa} of the surface tension  
$\sim 10^{-6} N/m$ (or $10^{-3}$ milliJoules per metre squared)
agrees in order of 
magnitude with $T/r^2$,
where $r \sim 80 nm$ is the radius of the tubule in the experiment.

We say a compact dimension is large if its radius
is much greater than the molecular scale $l_{mol}$. Let us
define $\xi=r/l_{mol}$.
After dimensional reduction, we find that the predicted
magnitude for the fluctuating line tension decreases by
a factor $1/\sqrt{\xi}$. This is in parallel with
Sorkin's observation\cite{large} that in models with large extra 
dimensions \cite{randall} the prediction of a fluctuating cosmological
constant results in a magnitude much smaller than the observed one.

{\it Conclusion:}        
We have developed an analogy between the surface tension
of membranes and the cosmological constant. The analogy is based
on the standard mapping between quantum field theory and statistical mechanics.
We have shown that the cosmological constant problem has its counterpart
in the context of membranes and suggested experimental probes for
measuring a fluctuating surface tension thus realising in analogy
Sorkin's proposal of a fluctuating cosmological constant.

The main new point of this paper is the {\it connection}
between two disparate fields. As some aspects are better understood
in one field and some in the other, we are able to derive insights
from {\it both} fields. For example, part b) is discussed in 
cosmology\cite{rafael}
but we have not seen a corresponding discussion in membranes.
The reverse is true for part a)\cite{safran}.
One could use the analogy to transport this discussion to cosmology.
One could invoke
an external reservoir to provide causet elements at Planck
density. Carrying over ideas from the membrane context, one
could introduce in analogy to $f(\alpha)$ a
``Quantum action per causet element'', a function of causet four-density
which has a minimum at the Planck four density. Both part a) and
part b) seem to emerge naturally from this description. We hope to 
interest the causet community in implementing this idea  in technical
detail.

The free energy of a membrane has a term proportional to the
intrinsic curvature, which in two dimensions is
purely topological. Such a term
can affect the global topology of the membrane. For example 
a large (negative) value for $a_2'$ favours the production of
handles. For instance, the ``Plumber's nightmare phase"
in membranes \cite{daoud} is an example of a phase where there is a
proliferation of handles.  
There are similar effects in the GR context too:
there are topological terms which one can add to the action 
at the next order (after the Einstein-Hilbert term). These
would be negligible at ordinary energies and length scales,
but at the Planck scale may favour the production of
handles, an idea which has been discussed before as ``space-time
foam''. 

The analogy we develop in this paper (like all analogies)
has its limitations. Obvious differences are those of dimension
(two versus four) and signature (Euclidean versus Lorentzian).
In the case of Lorentzian space-time the only distribution of causet
elements consistent with local Lorentz invariance is Poisson. In the case
of membranes, the distribution of amphiphiles on $\Sigma$ is certainly not
Poisson. There are correlations between the molecules. In spite
of this difference, $1/\sqrt{N}$ fluctuations of the surface tension
arise, due to the central limit theorem. Another important difference
is that in GR all the field variables which appear in the action
are intrinsic, whereas the membrane free energy has terms dependent
on the embedding.

One must remember that for finite size systems, the
choice of ensemble is crucial. For a discussion of this point 
and its operational realisation see \cite{ensemble}.
Throughout this paper we have worked in the constant area 
(or Helmholtz) ensemble: we keep the area of the membrane fixed and
study the surface tension fluctuations.
In the Helmholtz ensemble,
we expect to see fluctuations in the surface tension manifesting
as fluctuations in the bead about its mean position. 
In the conjugate
Gibbs ensemble where one holds the surface tension constant 
one would expect to
see area fluctuations. In analogy, in the cosmological
context, observers are viewing the universe at a particular
epoch or fixed $4$ volume ${\cal V}$ (or Helmholtz ensemble).  
There is no counterpart in cosmology
of the Gibbs ensemble, which would involve fixing the 
cosmological constant and letting the ``epoch'' fluctuate.

We have described a ``history'' (in analogy to a configuration)
as a timeless entity.
While this is correct in the mathematical analogy we work with,
a history is described in physical terms by its development or unfolding.
In causet theories, there are efforts to dynamically describe
causets in terms of elements coming into being one at a time.
The present description of a ``history'' as timeless
does not capture this unfolding of the causet dynamics.
In spite of its limitations,
we feel that the analogy we develop here is suggestive and
can be fruitfully pursued further.

{\it Acknowledgements:} We thank S. Surya for discussions on
causets and Y. Hatwalne for conversations on fluid membranes. 
We also thank R. Bandyopadhyay,  
D. Bhattacharya, R. Capovilla, D. Cho, J. Henson and  B. Nath for their 
comments. 
\newpage
\vspace{.6 cm}
\hspace{-3 cm}
\vbox{
\epsfxsize=18.0cm
\epsfysize=9.0cm
\epsffile{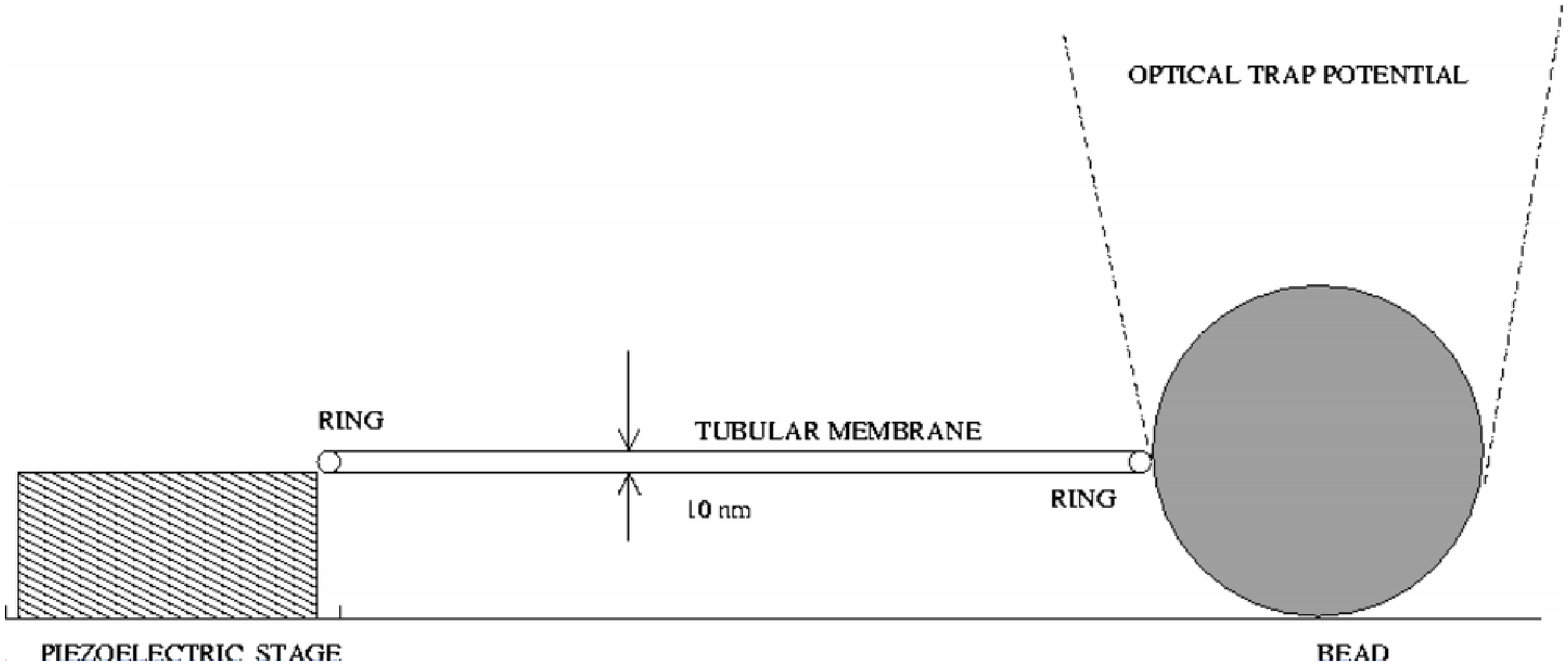}
\begin{figure}
\caption{Experimental Setup (not to scale) for measurement of the Surface 
Tension of a Fluid Membrane. The tubular membrane is stretched between 
two microscopic rings.}
\label{contrast}
\end{figure}}
\normalsize
\newpage
\normalsize
\centerline{\bf Table 1: The Analogy}
\begin{tabbing}
\noindent {\bf Membranes}~~~~~~~~~~~~~~~~~~     \= ~~~~~~~~~~~~\= 
{\bf 
Universe}
\\
\\
\noindent Configuration ${\cal C}$~~~    \> ~~~~\> History
${\cal H}$ 

\\
\\
\noindent Area of a configuration ${\cal A}$     \> ~~~~\> Four volume of 
a history ${\cal V}$

\\
\\
\noindent Sum over configurations~~~~~~     \> ~~~~\> Sum
over histories 

\\
\\

\noindent Energy ${\cal E(C)}$~~~~~~     \> ~~~~\> Classical Action ${\cal I(H)}$

\\
\\
\noindent Minimum energy configuration~~~~~~     \> ~~~~\> Classical Path 
Of Least Action

\\
\\
\noindent Temperature $T$ ~~~~~~     \> ~~~~\> Planck's constant $\hbar$

\\
\\
\noindent Thermal Fluctuations ~~~~~~     \> ~~~~\> Quantum Fluctuations

\\
\\
\noindent Surface Tension $\sigma$~~~~~~     \> ~~~~\> Cosmological 
Constant $\lambda$ 
\\
\\

\noindent Molecular Length $l_{mol}$     \> ~~~~\> Planck Length 
$l_{Planck}$ 

\\
\\
\noindent molecules     \> ~~~~\> Causet elements
\\
\\

\noindent Free Energy~~~~~~     \> ~~~~\> Quantum Action  
\\
\\
\noindent 
$E_0=a_0\int d^2 x \sqrt{\gamma} $ ~~~~~~~~~~~~~~~~    \>~~~~~~~~~~~~~\>  
$I_0=c_0\int d^4 x \sqrt{-g} $  
\\
\\
\noindent 
$E_2=\int d^2 x \sqrt{\gamma} H^2 $~~~~~~     \> ~~~~~~~~~~~~~~\> 
$I_2=c_2\int d^4 x 
\sqrt{-g} R$ 
\\
\\
\noindent {Molecular level discreteness of space}~~~~~~     \> 
~~~~~~~~~~~~~~~~~\> 
{Planck scale discreteness of spacetime}
\\
\\
\noindent{Plumber's nightmare phase}~~~~~~     \> 
~~~~\> 
{Spacetime foam}
\\
\\

\end{tabbing}
\newpage
\centerline{\bf Table 2: Typical Interfacial Tension Values}
\begin{tabbing}
\noindent {\bf Interfaces}~~~~~~~~~~~~~~~~~~     \= ~~~~~~~~~~~~\= 
{\bf 
Surface Tension } 
\\
\\
\noindent {}~~~~~~~~~~~~~~~~~~     \= ~~~~~~~~~~~~\= 
{\bf 
in milli Joules per metre squared} 
\\
\\
\noindent  Water-Vapour ~~~~~~     \> ~~~~\> 72.6  
\\
\\
\noindent Water-Oil~~~~~~     \> ~~~~\> 57  
\\
\\
\noindent Mercury-Water~~~~~~     \> ~~~~\> 415  
\\
\\
\noindent Glycerol-Air~~~~~~     \> ~~~~\> 63.4  
\\
\\
\noindent Decane-Air~~~~~~~~~     \> ~~~~~~~~\> 23.9  
\\
\\
\noindent Hexadecane-Air~~~~~~     \> ~~~~\> 27.6  
\\
\\
\noindent Octane-Air~~~~~~     \> ~~~~\> 21.8  
\\
\\
\noindent Water-Air~~~~~~     \> ~~~~\> 40  
\\
\\

\end{tabbing}


\begin{references}
\bibitem{gulliver} A similar problem faced the learned men
of Brobdingnag when confronted with Gulliver, who 
was tiny by the standards of Brobdingnag.
Why was he so small and why (if he was so small) did he exist
at all?
\bibitem{quint}
For instance ``quintessence models" suffer from the ``why now?''
problem \cite{rafael}. They need an energy scale of
$\hbar H_0 \sim 10^{-33} eV$ to turn on the field at the 
present epoch. $H_0$ here is the Hubble constant.
\bibitem{thickness}
In this case the molecular scale is set by the thickness of the
bilayer, which is related to the length of the surfactant molecules.
\bibitem{vesicle}
For example, surface tension fluctuations of micron sized vesicles
may be observable as shape fluctuations by light scattering.
\end{references}
\end{document}